\documentstyle[12pt,psfig]{article}

\begin{document}
\title{Current quark mass and g-2 of muon and  
$ee^+\rightarrow\pi^+\pi^-$}
\author{Bing An Li and Jian-Xiong Wang\footnote{On leave from Institute
of High
Energy Physics, Academia Sinica, 10039 Beijing, China} \\
Department of Physics and Astronomy, University of Kentucky\\
Lexington, KY 40506, USA}

\maketitle
\begin{abstract}
Based on a phenomenologically successful effective chiral theory of 
pseudoscalar, vector, and axial-vector mesons the dependences of 
$\rho-\omega$ mixing and the vertex $\omega\pi\pi$ on $m_d-m_u$
are found. Using the new data of 
$ee^+\rightarrow\pi^-\pi^+$, $m_d-m_u$ has been determined to be
4.24MeV. The form factor 
of pion agrees with data in both space- and time-like 
regions. The branching ratio of 
$\tau\rightarrow\pi\pi\nu$ is computed to be $25.14\%$ which 
agrees with data. CVC is satisfied.
The values of g-2 of muon from $\pi\pi$ channel
are calculated in the range of $q^2<1.3^2 GeV^2$. 
\end{abstract}

\pagebreak
In particle physics $ee^+\rightarrow\pi^-\pi^+$ is a very important process.
Recently, measurements of the cross section of $ee^+\rightarrow\pi^-\pi^+$
in the center of mass energy region from 0.61 to 0.96 GeV with a $0.6\%$
systematic uncertainty have been reported by CMD-2 group\cite{1}. In Ref.\cite{1}
two different models\cite{2,3} have been used to analyze the data. In this paper
we exploit another model, effective chiral theory of pseudoscalar, vector, and
axial-vector mesons\cite{4} to study $ee^+\rightarrow\pi^-\pi^+$.
The contribution of $\pi\pi$ channel to g-2 of muon has been calculated\cite{15}.
To get good accuracy the form factor of pion in both time-like, 
space-like, and $B(\tau\rightarrow\pi\pi\nu)$ have to be taken into
account\cite{15}.
However, in Ref.\cite{14} by using CVC it is obtained \(B(\tau\rightarrow\pi\pi\nu)
=(24.25\pm0.77)\%\). The data\cite{10} is $(25.40\pm0.14)\%$. The seperation of the 
isovector form factor of pion from $ee\rightarrow\pi\pi$ depends on the treatment 
of $\omega\rightarrow\pi\pi$. In this paper a different mechanism for $\omega
\rightarrow\pi\pi$ is provided.
The amplitude of $\omega\rightarrow\pi^-\pi^+$ is obtained from the 
effective theory\cite{4}. 
The mass difference of u and d quarks is determined by fitting the data\cite{1}. 
The contribution of $\pi\pi$
channel to g-2 of muon is calculated in the range of 
$q^2 < 1.3^2 GeV^2$. In these calculations both space-, time-like regions
and $\tau$ decays are taken into account.

Using the knowledge of current algebra and nonlinear $\sigma$ model,
the Lagrangian of the effective chiral theory of pseudoscalar, vector, 
and axial-vector mesons is constructed as\cite{4}
(taking two flavors as an example)
\begin{eqnarray}
{\cal L}=\bar{\psi}(x)(i\gamma\cdot\partial+\gamma\cdot v
+\gamma\cdot a\gamma_{5}
-mu(x))\psi(x)-\bar{\psi(x)}M\psi(x)\nonumber \\
+{1\over 2}m^{2}_{0}(\rho^{\mu}_{i}\rho_{\mu i}+
\omega^{\mu}\omega_{\mu}+a^{\mu}_{i}a_{\mu i}+f^{\mu}f_{\mu})
\end{eqnarray}
where \(a_{\mu}=\tau_{i}a^{i}_{\mu}+f_{\mu}\), \(v_{\mu}=\tau_{i}
\rho^{i}_{\mu}+\omega_{\mu}\),
and \(u=exp\{i\gamma_{5}(\tau_{i}\pi_{i}+
\eta)\}\), and M is the current quark matrix.
Since mesons are bound states solutions of $QCD$ they
are not independent degrees of freedom. Therefore,
there are no kinetic terms for meson fields. The kinetic terms
of meson fields are generated from quark loops.
m is a parameter, the constituent quark mass, which related to quark
condensate\cite{4}. The theory has dynamical chiral symmetry.
In the limit, $m_q\rightarrow 0$, the theory is explicitly chiral
symmetric. 

After integrating out the quark fields the 
effective Lagrangian of mesons is 
obtained and can be found in Ref.\cite{4}. Normalizing the kinetic terms of pion,
$\rho$, and $\omega$ fields, the fields of physical mesons are defined
\[\pi\rightarrow {2\over f_\pi}\pi,\;\;\;\rho\rightarrow{1\over g}\rho,\;\;\;
\omega\rightarrow{1\over g}\omega.\]
$f_\pi$ is the pion decay constant and g is the universal coupling constant.
They are defined as
\[f^2_\pi=F^2(1-{2c\over g})^{-1},\;\;\;
F^2=\frac{N_C}{(4\pi)^2}m^2\int d^4 k\frac{1}{(k^2+m^2)^2},\]
\[g^2=\frac{F^2}{6m^2},\;\;\;
c=\frac{f^2_\pi}{2gm^2_\rho}.\]
$f_\pi$ and g are two inputs. To define $F^2$ a cut-off has to be introduced\cite{4}.
 
Feynman diagrams at tree level are at $O(N_C)$ and loop diagrams
are at higher order in $N_C$ expansion. 
The theory is phenomenologically successful\cite{5}.

The Vector Meson Dominance(VMD) plays major role in studying the process
$ee^+\rightarrow\pi^-\pi^+$. VMD is 
a natural result of this theory and it has been derived as\cite{4}
\begin{eqnarray}
&&{1\over2}eg\{-\frac 12F^{\mu \nu }(\partial _\mu \rho _\nu
^0-\partial _\nu \rho _\mu ^0)+A^\mu j_\mu ^0\},\nonumber \\
&&{1\over6}eg\{-\frac 12F^{\mu \nu }(\partial _\mu \omega _\nu
-\partial _\nu \omega _\mu )+A^\mu j_\mu ^\omega\},\nonumber \\
&&-{1\over{3\sqrt{2}}}eg \{-\frac 12F^{\mu \nu }(\partial _\mu
\phi _\nu -\partial _\nu \phi _\mu )+A^\mu j_\mu ^\phi\},
\end{eqnarray}
The current $j_\mu^0$ is derived from the vertex of $\rho\pi\pi$\cite{4}
\begin{equation}
{\cal L}_{\rho \pi \pi }={2\over g}f_{\rho \pi \pi}(q^2)\epsilon _{ijk}\rho
_i^\mu \pi _j\partial _\mu \pi _k
\end{equation}
by substituting $\rho^0_\mu \rightarrow{1\over2}egA_\mu$
into this equation. In the same way we obtain
\begin{equation}
{\cal L}_{\gamma \pi \pi }=ieA_\mu f_{\rho \pi \pi}(q^2)
(\pi^+\partial_\mu\pi^- -\pi^-\partial_\mu\pi^+).
\end{equation}
where
\begin{equation}
f_{\rho \pi \pi }(q^2)= 1+\frac{q^2}{2\pi ^2f_\pi
^2}[(1-\frac{2c}g)^2-4\pi ^2c^2].
\end{equation}  
$q$ is the momentum of $\rho $ meson. 
$f_{\rho\pi\pi}(q^2)$ is an
intrinsic form factor which is the physical effect of quark loops. 
Similarly, $j^\omega_\mu$ and $j^\phi_\mu$ can be defined\cite{4} too.
We have used VMD and this effective theory to study the form factors of pion and
kaons\cite{6}, in which the $\rho-\omega$ mixing is taken to be a constant.
In this paper the amplitude of $\omega\rightarrow\pi\pi$ is determined by this 
effective theory\cite{4} completely.

$\omega\rightarrow\pi^-\pi^+$ contributes to $ee^+\rightarrow\pi^-\pi^+$.
The study of $\omega-\rho$ mixing has a long history. 
In this paper a different approach is taken to determine 
the contribution of $\omega\rightarrow\pi\pi$ to $ee\rightarrow\pi\pi$.
Based on the Lagrangian of mesons derived from Eq.(1)
$\omega-\rho$ mixing is found
\begin{equation}
{\cal L}_{\rho\omega}=\{-\frac{1}{4\pi^2g^2}{1\over m}(m_d-m_u)
+{1\over24}e^2 g^2\}(\partial_\mu\rho_\nu-\partial_\nu\rho_\mu)
(\partial_\mu\omega_\nu-\partial_\nu\omega_\mu).
\end{equation}
Eq.(6) has been presented in Refs.\cite{4,7}.
However, there is direct coupling between $\omega$ and $\pi\pi$
which is derived from Eq.(1) too
\begin{eqnarray}
\lefteqn{{\cal L}^d_{\omega\pi\pi}=\frac{2i}{g}f_{\omega\pi\pi}(q^2)
\omega_\mu(\pi^+\partial_{\mu}\pi^- 
-\pi^-\partial_{\mu}\pi^+)}\nonumber \\
&&+{i\over6}e^2 gf_{\rho\pi\pi}(q^2)\omega_\mu(\pi^+\partial_\mu\pi^-
-\pi^-\partial_\mu\pi^+),
\end{eqnarray}
where
\[f_{\omega\pi\pi}(q^2)=
-\frac{q^2}{2\pi^2 f^2_\pi}(1-{4c\over g})
{1\over m}(m_d-m_u).\]

Taking Eqs.(2,3,5,6,7) into account(Fig.1),
the
cross section of $ee^+ \rightarrow\pi^- \pi^+$ is obtained
\begin{eqnarray}
\sigma & = & \frac{\pi \alpha ^2}{3}\frac 1{q^2}(1-\frac{4m_{\pi
^{+}}^2}{q^2})^{{3\over2}}|F_\pi(q^2)|^2.
\end{eqnarray}
where $F_\pi$ is the pion form factor 
\begin{eqnarray}
&F_\pi(q^2)=f_{\rho\pi\pi}(q^2)\{1
-\frac{q^2}{q^2-m_\rho ^2 +i\sqrt{q^2}\Gamma _\rho (q^2)}
-{1\over3}\frac{b(q^2) q^2}{q^2-m_\omega ^2+i\sqrt{q^2}\Gamma _\omega
}\},\\
&{\displaystyle b(q^2)=-\frac{m_d-m_u}{m}\frac {f_{\omega\pi\pi}(q^2)}
{2 f_{\rho\pi\pi}(q^2)} 
   -\frac{(e g)^2}{12}  
   -(\frac{m_d-m_u}{2 \pi^2 g^2 m}-\frac{(e g)^2}{12})
    \frac{q^2}{q^2-m_\rho ^2 +i\sqrt{q^2}\Gamma _\rho (q^2)}}. \nonumber
\end{eqnarray}
Where $b(q^2)$ is the effect of $\omega\rightarrow\pi\pi$. 
In Eq.(9), the first term 
is from Fig.1(1), the second term from Fig.1(3), 
the last two
terms are from $\rho-\omega$ mixing(Fig.1(2,4,5)). 
 \(f_\pi=0.186GeV\) is taken in Ref.\cite{4}.
The decay width of
$\rho$ meson is calculated to be
\begin{eqnarray}
\Gamma _\rho (q^2) & = & \Gamma _{\rho ^0\rightarrow \pi ^{+}\pi
^{-}}(q^2)+\Gamma _{\rho ^0\rightarrow K\overline{K}}(q^2),
\nonumber \\ \Gamma _{\rho ^0\rightarrow \pi ^{+}\pi ^{-}}(q^2) &
= & \frac{f_{\rho \pi \pi }^2(q^2)\sqrt{q^2}}{12\pi g^2
}(1-\frac{4m_{\pi ^{+}}^2}{q^2})^{{3\over2}}\theta(q^2>4m_{\pi ^{+}}^2),
\nonumber
\\ \Gamma _{\rho ^0\rightarrow K\bar{K}}(q^2) & = & \frac{f_{\rho
\pi \pi }^2(q^2)\sqrt{q^2}}{48\pi g^2
}(1-\frac{4m_{K^{+}}^2}{q^2})^{{3\over2}}\theta(q^2>4m_{K^{+}}^2)
\nonumber \\ &&+\frac{f_{\rho \pi \pi }^2(q^2)\sqrt{q^2}}{48\pi
g^2}(1-\frac{4m_{K^0}^2}{q^2})^{{3\over2}}\theta (q^2>4m_{K^{0}}^2),
\end{eqnarray}
when $q^2>4m_{K}^2$ the $K\bar{K}$ channel is open. There are
other channels, however, in the range of $\sqrt{q^2}<1.3GeV$ the
contribution of other channels is negligible.
Because $\omega$ is a narrow resonance $\Gamma_\omega$ is taken to be
a constant. Following Ref.\cite{1} we take
\[m_\omega=(782.71\pm0.08)MeV,\]
\[\Gamma_\omega=(8.68\pm0.24)MeV.\]
Comparing with the two models of Ref.\cite{1}, there are two new effects:
\begin{enumerate}
\item Direct coupling of $\omega\pi\pi$(not through $\rho-\omega$
mixing),
\item the $\rho-\omega$ mixing depends on $q^2$.
\end{enumerate}

In Ref.\cite{4} \(m_\rho=0.77GeV\) and \(g=0.39\)\cite{5} are taken. By fitting
the new CMD2\cite{1} experiment data, we obtain
\begin{eqnarray}
&m_\rho=777.4\pm0.5MeV,\;\;\;g=0.3938\pm0.002, \nonumber \\
&m_d-m_u=4.24\pm0.32MeV.
\end{eqnarray}
The value of $m_d-m_u$ agrees with the one presented in Ref.\cite{8}. 
The fit is shown in Fig.2.
The comparison between the theory and experimental data of pion form
factor in both time-like(new and old) and space-like regions 
( $-1.1GeV<\sqrt{q^2}<1.3GeV$)(Fig.3). Theory agrees well with data.

The form factor of pion in space-like region is obtained by
taking $\Gamma_\rho$ and $\Gamma_\omega$ to be zero.  
The radius of pion is expressed as
\begin{equation}
<r^2_\pi>={6\over m^2_\rho}+{3\over2\pi^2 f^2_\pi}\{(1-{2c\over g})^2
-4\pi^2 c^2\}+\frac{e^2 g^2}{6m^2_\omega}=0.451 fm^2.
\end{equation}
The data\cite{9} is $(0.439\pm0.03)fm^2$.

Using Eqs.(6,7), we obtain
\begin{eqnarray}
\lefteqn{\Gamma(\omega\rightarrow\pi^+\pi^-)=|b(m^2_\omega)|^2
\Gamma_\rho(m^2_\omega)}\nonumber \\
&&B(\omega\rightarrow\pi\pi)=(0.97\pm0.19)\%.
\end{eqnarray}
The results of this paper and Ref.\cite{1} are listed in Tab.1.
\begin{table}
\begin{center}
\caption{Results}
\begin{tabular}{|c|c|c|c|} \hline
Parameter&GS model\cite{2}&HLS model\cite{3}&
This paper \\ \hline
$M_\rho(MeV)$&$776.09\pm0.64\pm0.50$&$775.23\pm0.61\pm0.50$&$777.4\pm0.50$
\\ \hline
$\Gamma_\rho(MeV)$&$144.46\pm1.33\pm0.80$&$143.88\pm1.44\pm0.80$&$147.37\pm0.40$
\\ \hline
$\Gamma(\rho\rightarrow e^+ e^-)$KeV&$6.86\pm0.11\pm0.05$&$6.84\pm0.12\pm0.05$
&$6.73\pm0.02$\\ \hline
$B(\omega\rightarrow \pi^+\pi^-)(\%)$&$1.33\pm0.24\pm0.05$&$1.43\pm0.24\pm0.05$
&$0.97\pm0.19$ \\ \hline
$\chi^2/\nu$&0.92&0.94&1.06\\ \hline
\end{tabular}
\end{center}
\end{table}
In this paper $m_\rho$ and g are inputs and all other quantities listed
in Tab.1 are theoretical values.
Tab.1 shows that the values of $m_\rho$, $\Gamma_\rho$, and 
$\Gamma(\rho\rightarrow ee)$ obtained in this paper are consistent with the 
values obtained by other two models. However, the central value of the 
branching ratio 
of $\omega\rightarrow\pi\pi$ obtained 
in this paper is smaller than others. The reason of this difference is 
how to treat the contribution of $\omega\rightarrow\pi\pi$. In Ref.\cite{1}
a mixing \(|\omega>=|\omega_0>+\epsilon |\rho_0>$ has been used, where
$\epsilon$ is a constant. In this paper 
the $\rho-\omega$ mixing obtained (Eq.6) strongly depends on $q^2$, 
and besides $\rho-\omega$ mixing
there is direct channel( Eq.7). It is quit clear here that 
the mechanism of $\rho-\omega$ mixing is the same as VMD(2).
 
Because of CVC the decay of $\tau\rightarrow \pi^+\pi^-\nu_\tau$
is determined by the isovector component of the pion form factor
\begin{equation}
\begin{array}{ll}
\displaystyle \frac{d\Gamma }{dq^2}&=\displaystyle \frac{G^2}{(2\pi )^3}\frac{|V_{ud}|^2}
{48m_\tau ^3}(m_\tau ^2+2q^2)(m_\tau ^2-q^2)^2\{1-\displaystyle \frac{4m_\pi
^2}{q^2}\}^{3/2}f^2_{\rho\pi\pi}(q^2)
\displaystyle \frac{m^4_\rho+q^2\Gamma^2_\rho(q^2)}{(q^2-m^2_\rho)^2+q^2\Gamma^2_\rho(q^2)}.
\end{array}
\end{equation}
The branching ratio is computed to be
\begin{equation}
B(\tau\rightarrow\pi^0\pi^-\nu)=(25.14\pm0.31)\%.
\end{equation}
The data is $(25.40\pm0.14)\%$\cite{10}. Theory agrees with data. CVC is satisfied.
The reason of the satisfaction of CVC in this paper is the new amplitude of
$\omega\rightarrow\pi\pi$(9). We predict a small $B(\omega\rightarrow\pi\pi)$
and obtain correct $B(\tau\rightarrow\pi\pi\nu)$.
 
Now we are ready to calculate the contribution of $\pi\pi$ 
channel to g-2 of muon. The results are
\begin{equation}
\begin{array}{ll}
&a^{(2)}(h.v.p)=\displaystyle{\int^\infty_{4m_\pi^2}dt {1 \over 4 \pi^3}
\sigma(t)k(t)},\\
&k(t)=\displaystyle{\int^1_0 dx {x^2(1-x) \over x^2+(1-x)t/m_u^2}}, \\
&a^{(2)}(2\pi,t<0.8GeV^2)=(4697\pm 34)\times 10^{-11}, \\
&a^{(2)}(2\pi,t<1.3^2GeV^2)=(4971\pm 47)\times 10^{-11}, \\
&a^{(2)}(2\pi,(0.61)^2<t<(0.96)^2 GeV^2)=(3721\pm 40)\times 10^{-11}, \\
\end{array}
\end{equation}
where $t=q^2$. 

It is necessary to point out that the results are based on an effective 
chiral theory. 
However, in the energy region of Ref.\cite{1} our result agrees with Ref.\cite{1} 
within their errors, $(3681\pm26\pm22)\times 10^{-11}$. 
Theoretical results fit the form factors in both 
space-like and time-like regions and $\tau$ decays. 
CVC is satisfied. On the other hand, 
this effective theory has been applied to study many other physical processes.
Theory agrees with data well. 
We will present the results of many other channels in
the near future. 

In summary, the CMD2 new data has been used to determine the value of $m_d-m_u$. 
A smaller $B(\omega\rightarrow\pi\pi)$ is obtained. 
The form factor of pion agrees with data in both space-like and time-like regions.
Theoretical $B(\tau\rightarrow\pi\pi\nu)$ agrees with data and CVC is satisfied.
The contribution of $\pi\pi$ channel to g-2
of muon at $\sqrt{q^2}<1.3GeV$ has been calculated. 
In this study all vertices are derived up to the fourth order in derivatives.
They fit the data up to $1.3^2 GeV^2$.
To fit the data at higher energies we need to add terms at higher order in 
derivatives and hard gluon effects. We will present the study in the near future.

The study is supported by DOE grant No.DE-91ER75661.

\pagebreak

\pagebreak
\begin{flushleft}
{\bf Figure Captions}
\end{flushleft}
{\bf FIG. 1.} Feynman Diagrams of $e^{+}e^{-}\rightarrow \pi ^{+}\pi ^{-}$ \\
{\bf FIG. 2.} Charged Pion form factor fitting by using our theoretical
formula and the data are from Ref.\cite{12}. $X=\sqrt{q^2}$\\
{\bf FIG. 3.} Charged Pion form factor in time-like region and space-like region.
Data are from \cite{1,9,11,12,13}. 
$X=\sqrt{q^2}\,\, when\,\, q^2>0$ and $X=-\sqrt{-q^2}\,\,when\,\, q^2<0$\\

\begin{figure}
\begin{center}
\psfig{figure=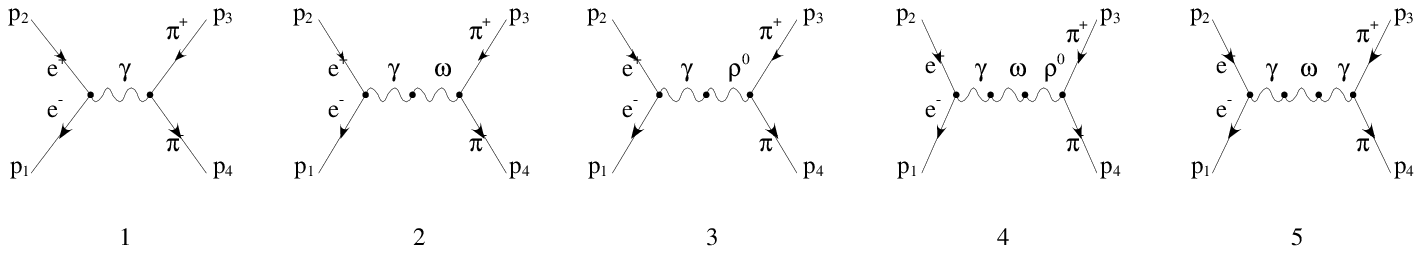}
FIG. 1.
\end{center}
\end{figure}

\begin{figure}
\begin{center}
\psfig{figure=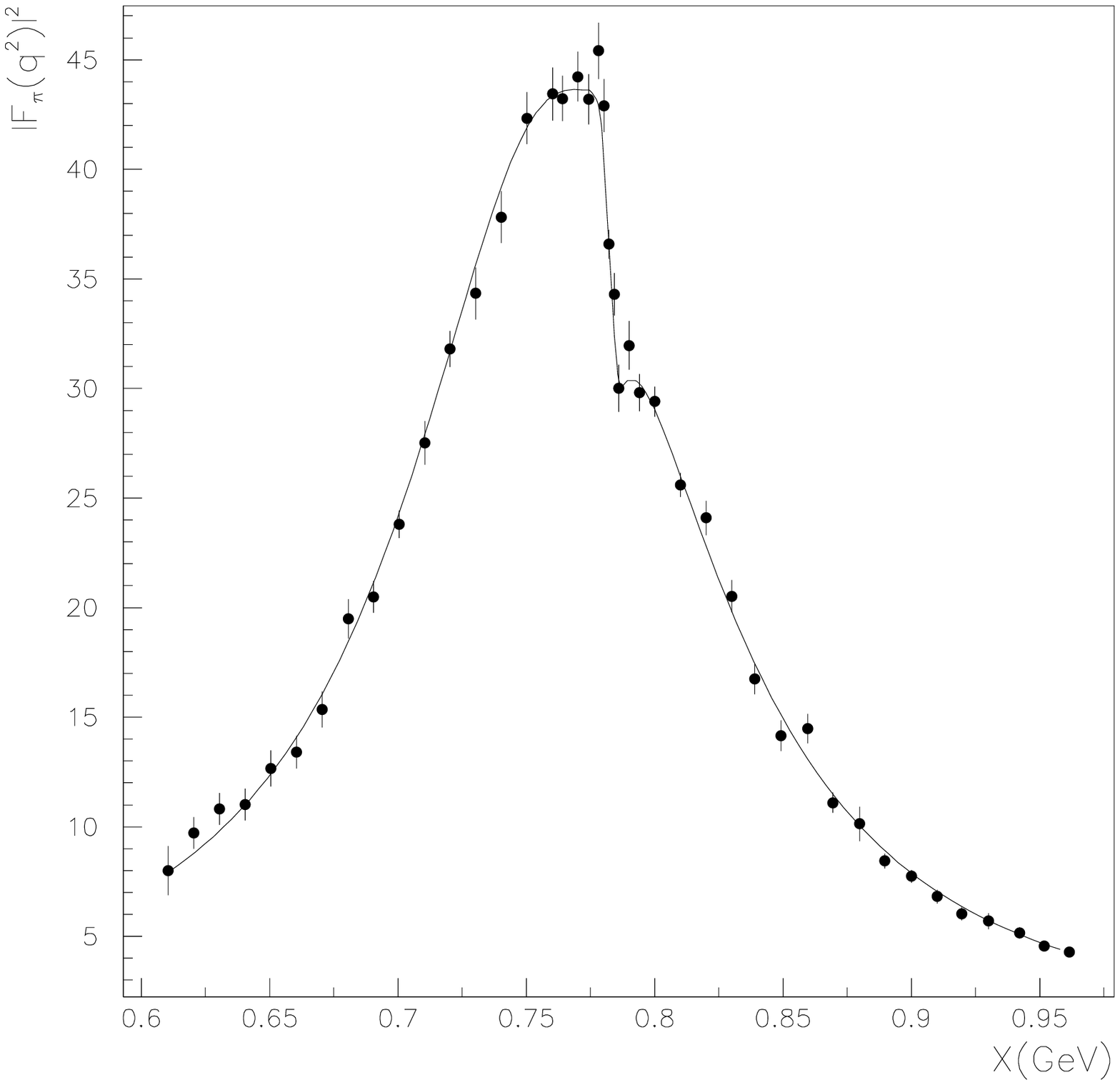}
FIG. 2.
\end{center}
\end{figure}

\begin{figure}
\begin{center}
\psfig{figure=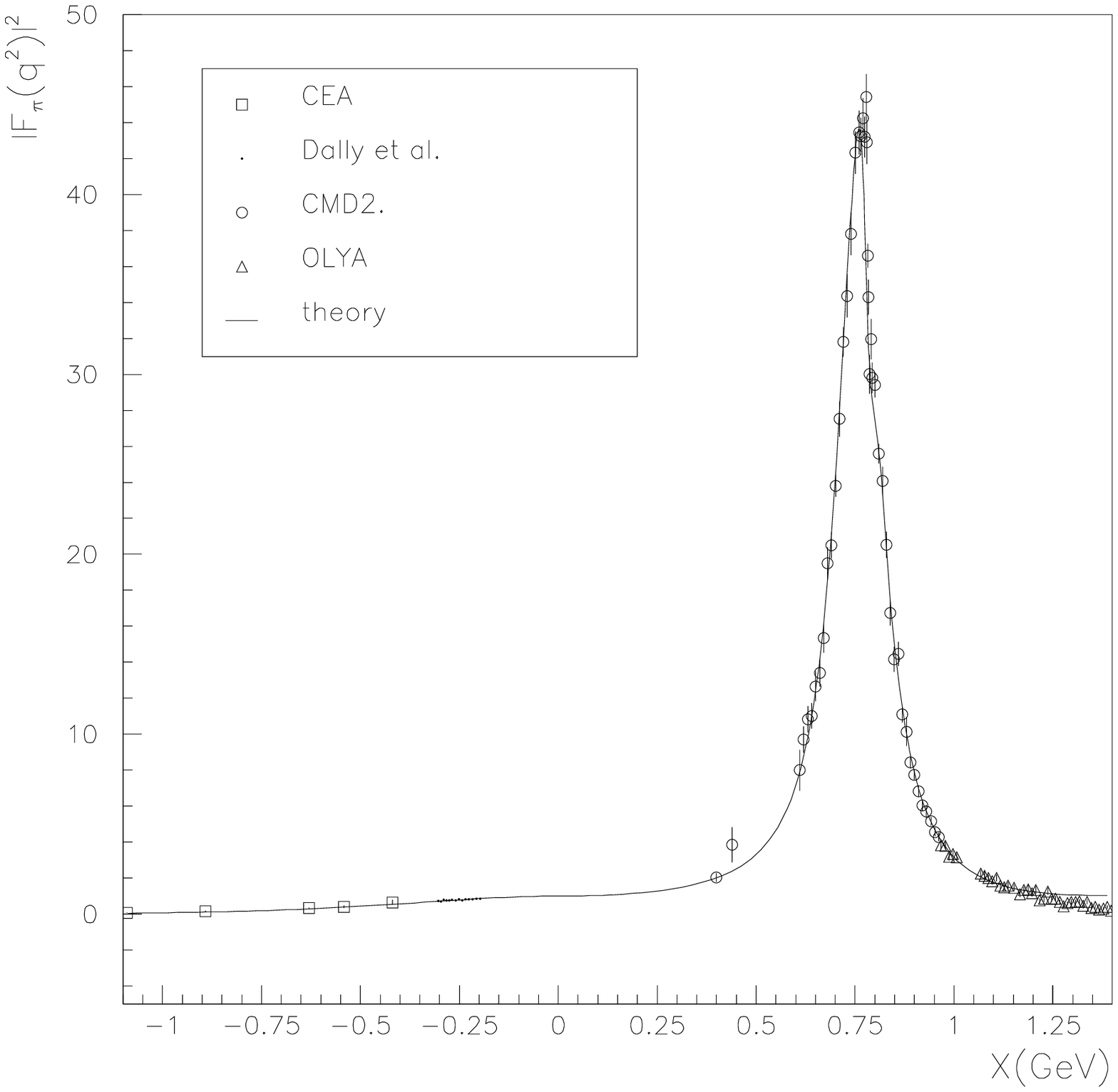}
FIG. 3.
\end{center}
\end{figure}


\begin{thebibliography}{30}
\bibitem{1} R.R. Akhmetshin at al. Phys. Lett. B527, 161(2002)
\bibitem{2} G.J.Gounaris and J.J.Sakurai, Phys. Rev. Lett.,{\bf 21},244(1968).
\bibitem{3} M.Benayoun at al., Eur.Phys.J.,{\bf c2},269(1998);
M.Bando et al., Phys.Rev.Lett., {\bf 54},1215(1985).
\bibitem{4} B.A. Li, Phys. Rev. D {\bf 52}, 5165(1995); {\bf 52}, 5184 (1995).
\bibitem{15} see review article by W.J.Marciano and B.L.Roberts, hep-ph/0105056.
\bibitem{14} E.I.Eidelman and V.N.Ivanchenko, Nucl. Phys., {\bf B}(Proc. Suppl.)
{bf 55C},181(1997).
\bibitem{10} Particle Data Group, Eur.Phys.J, {\bf C15}, 1-878(2000).
\bibitem{5} B.A. Li, Phys. Rev. D {\bf 55}, 1436 (1997); {\bf 55} 1425 (1997);
D.N. Gao, B.A. Li, and M.L. Yan, Phys. Rev. D {\bf 56}, 4115 (1997);
B.A. Li, D.N. Gao, and M.L. Yan, Phys. Rev. D {\bf 58}, 094031
(1998); T.L.Zhuang, X.J.Wang, and M.L.Yan, Phys.Rev., {\bf D62},053007,2000.
\bibitem{6} J.Gao and Bing An Li, Phys.Rev., {\bf D61},113006,20000.
\bibitem{7} D.N.Gao and M.L.Yan, Eur.Phys.J, {\bf A3},293(1998).
\bibitem{8} H.Leutwyler, hep-ph/9609467.
\bibitem{9} E.B. Dally et al., Phys. Rev. Lett. {\bf 48}, 375 (1982).
\bibitem{12} C. N. Brown et al., Phys. Rev. D8, 92(1973),
\bibitem{11} A. Quenzer et al., Phys. Lett. {\bf 76B}, 512 (1978); L.M.
Barkov et al., Nucl. Phys. {\bf 256B}, 365 (1985).
\bibitem{13} C.J. Bebek et al., Phys. Rev. D {\bf 13}, 25 (1976).
\end{thebibliography}
\end{document}